\documentclass[iconfrence]{IEEEtran}

\usepackage{url}
\usepackage{cite}
\usepackage{graphicx}
\usepackage{amsmath,amsfonts,amssymb,subfigure}

\begin{document}

\title{ECN verbose mode: a statistical method for network path congestion estimation}

\author{
\IEEEauthorblockN{R\'emi Diana$^{1,3}$,  Emmanuel Lochin$^{2,3}$}
\IEEEauthorblockA{\\$^1$T\'eSA/CNES/Thales, Toulouse, France\\
$^2$CNRS ; LAAS ; 7 avenue du colonel Roche, F-31077 Toulouse, France\\
$^3$ Universit\'{e} de Toulouse ; UPS, INSA, INP, ISAE ; LAAS ; F-31077 Toulouse, France
}
}

\maketitle

\begin{abstract}
This article introduces a simple and effective methodology to determine the level of congestion in a network with an ECN-like marking scheme. The purpose of the ECN bit is to notify 
TCP sources of an imminent congestion in order to react before losses occur. However, ECN is a binary indicator which does not reflect the congestion level (i.e. the percentage of queued packets) of the bottleneck, thus preventing any adapted reaction. In this study, we use a counter in place of the traditional ECN marking scheme to assess the number of times a packet has crossed a congested router. Thanks to this simple counter, we drive a statistical analysis to accurately estimate the congestion level of each router on a network path.
We detail in this paper an analytical method validated by some preliminary simulations which demonstrate the feasibility and the accuracy of the concept proposed. We conclude this paper with possible applications and expected 
future work.
\end{abstract}

\begin{keywords}
Congestion estimation, ECN, measurements
\end{keywords} 

\section{Introduction}

While dropping packets to prevent congestion was considered as a paradox, many studies have shown the undeniable assets of the Explicit Congestion Notification flag \cite{rfc2481}. The story starts in 1994 when Sally Floyd shows that this notification allows to increase TCP performances \cite{floyd94ecn} and later in \cite{le03effect}, where the authors reach similar conclusion concerning the web traffic. At last, Aleksandar Kuzmanovic in ``The Power of Explicit Congestion Notification'' \cite{power} investigates the pertinence of ECN and demonstrates once again, that ECN's users will obtain better performances even if all the Internet is not fully ECN-capable.

The following study \cite{medina05measuring} published in 2004 precises that ECN is only used by 2,1\% of computers and that this low percentage can be partly explained by firewall, NAT and other \emph{middle-boxes} of the Internet which reset (without any justification) the ECN flag. However, this is definitely not the main reason. Indeed, although this flag is currently implemented both in end-hosts (GNU/Linux, Mac OSX and Windows Vista) and inside the core network (Cisco IOS implements a RED/ECN variant called WRED/ECN), ECN remains surprisingly disabled by default for all these systems. Concerning end-hosts, this might appear paradoxical. While today CUBIC and Compound TCP variants are enabled by default (respectively in GNU/Linux and Windows Vista) and are still under debate concerning their friendliness with the current Newreno TCP version, a proved mechanism as ECN is not. 

We believe this trend has two main reasons: firstly, this is partly due to the behaviour of TCP face to ECN marked packets. Indeed, the goal of the ECN bit is to notify TCP sources of an imminent congestion but this binary indicator does not reflect the real network congestion level. Intuitively, CUBIC and Westwood protocols might better perform than TCP Newreno/ECN due to the nature of the information returned by the ECN binary signal which does not provide any quantitative estimation of the congestion level allowing TCP to efficiently adapt its sending rate\footnote{We remark that there is a lack of performances evaluation study between ECN-compliant protocols and new proposals such as CUBIC for instance. At least, a recent study clearly shows a clear disequilibrium between TCP Newreno and CUBIC \cite{stewart09}.}. In other words, whatever the number of ECN marked, the TCP reaction is to halve the congestion window and this action is not well adapted to all cases. Secondly, CUBIC and Westwood are pure end-to-end solutions and as a result, are much more easier to deploy while TCP/ECN must involve both the core network and the end-hosts.
However, several research work demonstrate that the design of a mechanism to optimally manage network congestion and capacity while being fair with other flows cannot be done without network collaboration \cite{bmcc,rfc3649,xcp}. Unfortunately and to the best of our knowledge, the major barrier is that we do not have today a solution, that do not involve complex computation inside the core routers (such as BMCC \cite{bmcc} or XCP \cite{xcp}), able to assess at the sender side the exact congestion level of the bottleneck of the path allowing a transport protocol such as TCP to correctly react to this congestion. For instance, BMCC introduces complex mechanisms inside the router and is only compliant with IPv4 (due to the use of the 16 bits \texttt{IPid} field of the IPv4 header) while XCP involves large architectural changes.

This fact motivates the present study which proposes a statistical algorithm to assess the congestion level at the end-hosts side (i.e. receiver or sender sides) without
involving complex computation inside the core network. In particular, we aim at providing a practical solution to return concrete congestion measurements to the sender 
in order to avoid blind, approximate or excessive reaction from the source. The only modification deals with the marking method which is changed from a binary field to a count field similar to the TTL field from the IP packet. Practically, we do not have to extend the IP headers as the DiffServ Codepoint field is large enough to enable our proposal. We could argue, as in \cite{bmcc}, whether such modification involves or not heavy IETF standardization process, however we claim that it would be much more complex and uncertain to convince networking companies to add complex estimation method inside their own routers. Furthermore, this solution is generic enough to consider, as for ECN, this flag either as a simple binary indicator or as a counter. Finally, we point out that a recent IETF group named ConEx (Congestion Exposure) \cite{conex}, attempts to enable congestion to be exposed within the network layer of the Internet. The main candidate solution is to date re-ECN \cite{re-ecn} and propose the use of a second bit inside the IP header in order to differentiate the congestion upstream and downstream from an observation point inside the network. Internet service providers are pushing this idea as this would provide an essential tool (currently missing) to better manage and control their traffic\footnote{See the IETF \texttt{[re-ecn]} mailing-list and \cite{conex} for further details.}. If this solution is adopted, we could assist to a larger deployment of the ECN 
field that would facilitates the deployment of our proposal.

Following this new marking scheme, we propose a simple method 
which permits an accurate estimation of the congestion level experienced inside the routers of a given path. We present in a first part the mathematical basis of our proposition then, we develop in a second part our simulations and the practical analysis to evaluate the congestion level. Finally, we conclude about the possibility offered by this solutions and detail the remaining work.

\section{Marking proposal}

The ECN bit, as defined in RFC 3168, is a binary field of the IP header. This field can only contain a boolean value which informs a sender whether if a packet has crossed at least one congested router. Thus, it is impossible to distinguish a packet marked one time from those marked several times and which would have crossed several congested routers. This prevents any accurate metrology analysis of the link observed for the sake, for instance, of an adapted reaction from the source. In fact, a packet ECN-capable crossing a link composed by two routers and respectively marking at $30\%$ and $40\%$ will have a probability to be marked of $58\%$ (\emph{i.e.} $1-(1-0.4)(1-0.3)$). Obviously, this does not reflect the level of congestion of the network bottleneck (in this example: $40\%$) and could lead to an excessive reaction from the source. Thus, we propose to enhance the information returned with an incremental field (denoted ECN*) to count how many times a packet is marked.
The marking scheme, as for RED/ECN, strictly follows the RED algorithm \cite{red}. We will use this new metric (\emph{i.e.} how many times a packet is marked) to determine the level of congestion of the bottleneck. A RED/ECN* router will increment this counter instead of simply setting the ECN field to one. 
Through the analysis of the data received, a source can build the distribution of the marked packets.
Obviously, we cannot use this metric as it stands, in the following, we present the analytical method to interpret the data collected.  

\section{Analytical Study}

We present in this part the statistical analysis allowing us to process the data collected with our marking proposal. The results obtained allow to establish a relationship between the frequency of ECN* marked packets and the queue size of routers of the path.

\subsection{Hypothesis and notations}

We consider a topology of $n$ core routers in a row. For $1\leqslant i\leqslant n$ , we note R$_i$ the router number $i$. All these $n$ routers adopt the previously exposed ECN* marking scheme. Each router drops packets only if its queue is full. We consider that the congestion inside the network is stable. 
Thus, this relative network stability induces a constant congestion level and as a result, a constant average queue size for each router. Moreover, we know that a router decides to mark a packet only by analyzing its average queue size, thus we expect to obtain a constant marking probability for each routers. We call ``marking rate'' this probability. 		
In the rest of this paper, we adopt the following notations: 

\begin{itemize}
\item $n$: number of congested routers;
\item $p_{i}$: marking rate of the $i^{th}$ router from a path of $n$ routers;
\item $M_{k}^{n}$: a packet is marked $k$ times;
\item $p(M_{k}^{n})$: the probability of the event M$_{k}^{n}$;
\item $\sigma_{k}^{n}$: the $k^{th}$ elementary symmetric polynomial with $n$ variables. We remind: 
$\sigma_{k}^{n} = \sum_{1 \leqslant j_1 < j_2 < \cdots < j_k \leqslant n} x_{j_1} \cdots x_{j_k} $
\end{itemize}

\subsection{A first simple example : case of two routers}
\label{deuxrouteurs}

Let's assume a topology of two congested core routers $R_1$ and $R_2$ ($n=2$). In this example, we want to determine the marking rate of both routers with data collected by the sender positioned before $R_1$. In the same way as standard ECN which uses an ECN echo, the value of the counter ECN* is sent back to the sender with the TCP acknowledgement. Following the previous notations, we call $p_1$ and $p_2$ the marking rate of respectively $R_1$ and $R_2$. A simple calculation shows that a connection will observe a packet marked with a probability of $1 - (1 - p_1)(1 - p_2)$. 
Thus, with a standard ECN field, the sender cannot differentiate the two marking rates and so interprets a global congestion which is higher and not representative of the real congestion state. 
With our proposition ECN*, we refine this information sent back to the sender thanks to the determination of the marking rate of each crossed router. Thus, the sender can determine the level of the bottleneck queue and so could react in a more adapted way to the congestion state. In this example, we can estimate the ratio not only of the marked packets but also of packets marked one and two times. The sender can now estimate $p(M_1^2)$ and $p(M_2^2)$. These values become the new entries of the problem. If we develop these probabilities we have: 

	\begin{eqnarray*}
		p(M_{1}^{2}) & = & p_{1}(1-p_{2}) + p_{2}(1-p_{1}) = \sigma_{1}^{2} - 2 \sigma_{2}^{2}\\
		p(M_{2}^{2}) & = & p_{1} p_{2} = \sigma_{2}^{2}
	\end{eqnarray*} 
which is equivalent to: 			
	\begin{eqnarray*}
		p(M_{1}^{2}) & = & \begin{pmatrix} 2 \\ 0 \end{pmatrix} \sigma_{1}^{2} - \begin{pmatrix} 2 \\ 1 \end{pmatrix}  \sigma_{2}^{2}\\
		p(M_{2}^{2}) & = & \begin{pmatrix} 2 \\ 0 \end{pmatrix} \sigma_{2}^{2}
	\end{eqnarray*} 
	
Thanks to these equations, the sender can easily determine $\sigma_{1}^{2}$ and $\sigma_{2}^{2}$. Thus, using the existing relationship between the polynomial coefficients and the elementary symmetric function of its roots, the sender can evaluate $p_1$ and $p_2$ (here, $p_1$ and $p_2$ are the roots of the polynomial $P(x) = x^2$ - $\sigma_{1}^{2}$ x + $\sigma_{2}^{2}$). We detail in the following part how to compute in a more general way the polynomial to find the different $p_k$. Of course, the sender cannot associate each marking rate with the corresponding router but it gets a correct estimation of the congestion level of the bottleneck.		

We develop this case 
as it constitutes the basis of the proof by mathematical induction for the general formula of $p(M_{k}^{n})$. Indeed, when the distribution of the marked packets is done, the crucial step is the deduction of the $\sigma_{k}^{n}$. To do this, we use the formula of $p(M_{k}^{n})$ and a basic system resolution. Then, as shown in the following part, the determination of the polynomial roots give us the different $p_i$. 

The general formula has the following form: (this formula is demonstrated in Appendix)
	\begin{equation}
	 \forall k, \ 1\leqslant k\leqslant n, \ \ \ p(M_{k}^{n}) = \sum_{i=0}^{n-k} (-1)^{i} \begin{pmatrix} i+k \\ i \end{pmatrix} \sigma_{i+k}^{n}
	\label{eq:generale}
	\end{equation}

\subsection{Resolution}

Since the formula is now established, we now have to detail the operations a sender has to realize in order to deduce all the marking rates of the congested routers of its path. We detail and recall in this part the different steps mandatory to obtain the result. First of all, thanks to the distribution of the marked packets, the sender can estimate all the $p(M_{k}^{n})$. Indeed, the $p(M_{k}^{n})$ value is only the ratio between the number of packets marked $k$ times and the total number of received packets by the sender. Moreover, using (\ref{eq:generale}), the sender can compute the $\sigma_{k}^{n}$. Indeed, if we develop these relations we obtain : 

	\begin{align*}
		p(M_{1}^{n}) &=\sigma_{1}^{n} - 2 \sigma_{2}^{n} + \cdots + (-1)^{n-2} \begin{pmatrix} n-1 \\ n-2 \end{pmatrix} \sigma_{n-1}^{n} \\
				& \ \ \ \ \ \ \ \ \ \ + (-1)^{n-1} \begin{pmatrix} n \\ n-1 \end{pmatrix} \sigma_{n}^{n}  \\
		p(M_{2}^{n}) & =  \sigma_{2}^{n} - 3 \sigma_{3}^{n} + \cdots + (-1)^{n-2} \begin{pmatrix} n \\ n-2 \end{pmatrix}\sigma_{n}^{n} \\
		\vdots \ \ \ \  & =  \ \ \ \ \ \ \ \vdots   \\
		p(M_{n-2}^{n}) & =  \sigma_{n-2}^{n} - (n-1) \sigma_{n-1}^{n} + \begin{pmatrix} n \\ 2 \end{pmatrix}\sigma_{n}^{n} \\
		p(M_{n-1}^{n}) & =  \sigma_{n-1}^{n} - (n) \sigma_{n}^{n} \\
		p(M_{n}^{n}) & = \sigma_{n}^{n} 
	\end{align*}	
	\\
The unknowns are the $\sigma_{k}^{n}$, we obtain a diagonal system with $n$ equations and $n$ unknowns. The resolution is trivial.

\subsubsection{The Solving Polynomial}
As the previous system is solved, all the $\sigma_{k}^{n}$ are known. We now have to deduce the $p_{i}$. As said previously, the $\sigma_{k}^{n}$ are elementary symmetric functions. Thus, using the relationship between a polynomial and the elementary symmetric functions of its roots we can deduce the $p_{i}$. We detail this step in the following.
Let be $P(x)$ a polynomial of degree $n$, we write $P(x)$ as follows:  

	\begin{equation}
		P(x) = \sum_{m=0}^{n}{a_{m}x^{m}} 
		\label{eq4}
	\end{equation}
		
Let be $p_{i}$, ${1\leqslant i\leqslant n}$ the $n$ roots of $P$. Thus:
 
	$$\forall k, \ 1\leqslant k\leqslant n,~\sigma_{k}^{n} = \sum_{1 \leqslant j_1 < j_2 < \cdots < j_k \leqslant n} p_{j_1} \cdots p_{j_k} $$

	Moreover, we have the following relationships: 

	\begin{equation}
		\forall k, \ 1\leqslant k\leqslant n,~\sigma_{k}^{n} = (-1)^k \frac{a_{n-k}}{a_{n}}
		\label{eq5}
	\end{equation}
	
	We set 	$a_{n}$ = 1 in (\ref{eq5}). Then (\ref{eq4}) becomes: 	
		
	\begin{equation}
		P(x) = \sum_{k=0}^{n}{(-1)^{n-k} \sigma_{n-k}^{n}x^{k}} \nonumber
	\end{equation}
		
	So, we have a $n$ degree polynomial where the roots correspond to the $n$ marking rates of the $n$ crossed congested routers of the path. We just need now to estimate these roots.	
		
	\section{Simulation}
In this section, we evaluate our algorithm with data obtained with an ns-2 simulation. This section is divided in three points. First, we present the topology used in the ns-2 simulation and the results. Then, we present the establishment of the solving polynomial and a subtlety for its resolution. Finally, we present our results, a comparison with expected results and a brief discussion about these two last points.
		
	\subsection{Tests Topology and gathering of data} 

\begin{figure}[h] 
		\begin{center}	
		\includegraphics[width=6cm]{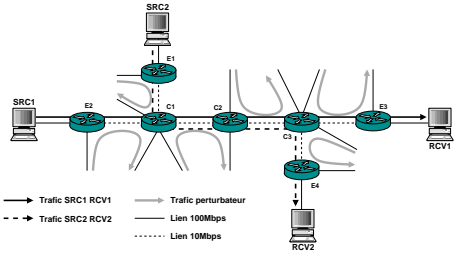} 
		\caption{Topology used for the simulation} 
		\label{topo_simu} 
		\end{center}
	\end{figure}

The topology used for the tests is given Figure \ref{topo_simu}. We use TCP/Newreno flows and the reaction of the senders to ECN is disabled. As a result, they do not react with a decrease of their congestion window when they receive an ECN marked acknowledgement. We have implemented our ECN* field and all the RED/ECN* routers use the same parameters: $min_{th}=50$, $max_{th}=100$, $max_p=1$ with a queue length of $100$. Concerning the disturbing flows aggregate, an accurate tuning of the senders' emission window has been necessary to simulate a distributed congestion.
The analysis of the data is done after 10 minutes when we consider the network stable (this corresponds to a generation of $50000$ packets). We analyze the two following TCP flows: the flow \#1 from SRC1 to RCV1 and the flow \#2 from SRC2 to RCV2. The topology voluntary presents two routers in common to estimate the impact of crossed traffics on our algorithm.

\begin{figure}[htb!] 
\subfigure[Results for flow \#1]{\includegraphics[width=4.2cm]{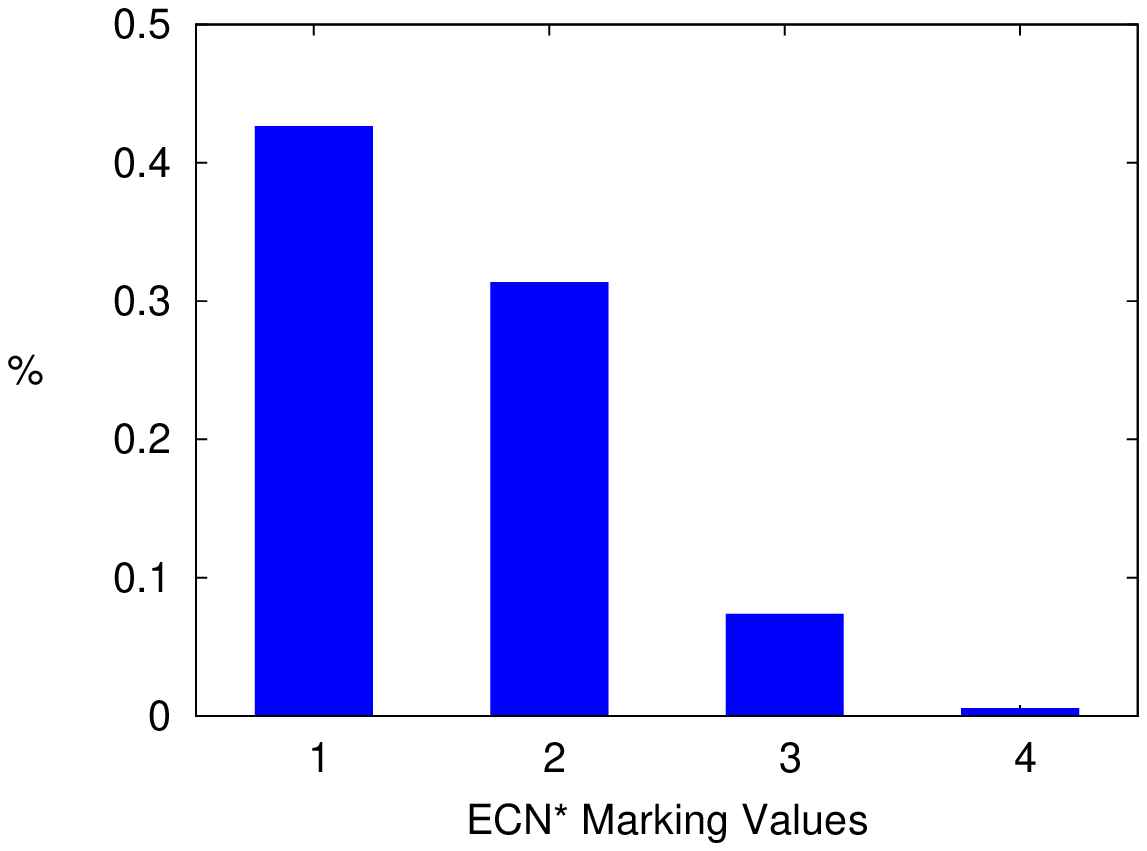}\label{graphEr1}}
\subfigure[Results for flow \#2]{\includegraphics[width=4.2cm]{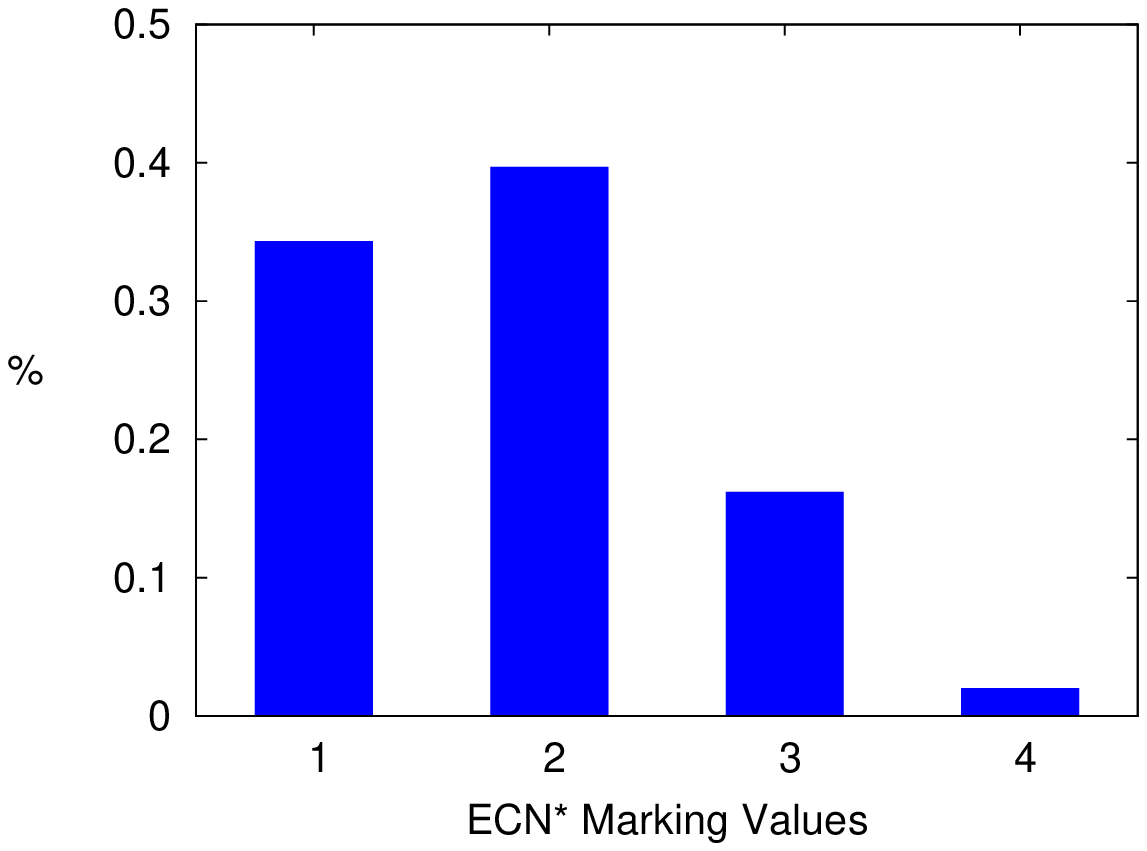}\label{graphFr1}}
\caption{Distribution of ECN* marked packets}
\end{figure}

The statistic study consists in building the histogram of the distribution of the values of the ECN* marking field for the flows \#1 and \#2. 
These results are presented in Figures \ref{graphEr1} and \ref{graphFr1}.

 \subsection{Determination of solving polynomial for flow \#1}
	
Figure \ref{graphEr1} gives the following results (here $n = 4$):				

	$$\left\lbrace
		\begin{array}{lll}
		p(M_{1}^{4}) & = 0.4264  & = \sigma_{1}^{4} - 2 \sigma_{2}^{4} + 3 \sigma_{3}^{4} - 4 \sigma_{4}^{4}\\
		p(M_{2}^{4}) & = 0.3134  & = \sigma_{2}^{4} - 3 \sigma_{3}^{4} + 6 \sigma_{4}^{4}\\
		p(M_{3}^{4}) & = 0.0738  & = \sigma_{3}^{4} - 4 \sigma_{4}^{4}\\
		p(M_{4}^{4}) & = 0.00548 & = \sigma_{4}^{4}
		\end{array}
	\right.$$
	
We then deduce the following $\sigma_{k}^{4}$: 	

	$$\left\lbrace
		\begin{array}{l}
		\sigma_{1}^{4} = 1.297 \\
		\sigma_{2}^{4} = 0.5676 \\
		\sigma_{3}^{4} = 0.0957 \\
		\sigma_{4}^{4} = 0.00548
		\end{array}
	\right.$$

By applying the method previously described we have: 
	$$ P(x) = x^{4} - 1.297 x^{3} + 0.5676 x^{2} - 0.0957 x + 0.00548  $$
				
\subsection{Practical Resolution}
\label{reso}
	
As the solving polynomial is built, we now have to solve $P(x) = 0$. The four roots of $P(x)$ correspond to the four marking rates of the four congested routers crossed by packets arriving to RCV1. As this problem is a stochastic one, we have to consider an uncertainty on the measurements obtained with the simulations. Indeed, unless having an infinite number of packets, we have to consider a drift. We take this possible drift in consideration in the determination of roots of $P(x)$. Basically, we resolve $P(x) \ = \ \epsilon \text{~for~} -10^{-3}\leqslant \epsilon \leqslant ~10^{-3}$. Thus, we obtain four ``areas of roots'' instead of ``solving roots''. We consider that the good value as the middle one. We note $\epsilon_{min}$ and $\epsilon_{max}$ the extreme values of $\epsilon$ from which $P(x) = \epsilon$ have four solutions. Indeed, if we have packets marked four times, we have to determine four solutions of the equation $P(x) = \epsilon$. This condition allows us to determine these four areas of roots.

In our example, we obtain the four following areas of roots: $[0.075 , 0.14]$   $[0.14 , 0.28]$   $[0.34 , 0.50]$   $[0.52 , 0.57]$. This allows us to deduce the four following marking rates : $11\%$, $21\%$, $42\%$ and $55\%$. With the same reasoning, we obtain for the flow \#2 the four following root areas : $[0.17 , 0.23]$   $[0.24 , 0.38]$   $[0.40 , 0.49]$   $[0.73 , 0.74]$ and so the four following marking rates : $20\%$, $31\%$, $44\%$ and $74\%$.
These results are presented in the Tab \ref{tabcomplet}.

		\subsection{Results interpretation}

We now compare the results computed with the average queue length of each RED/ECN* routers measured during the simulation. Thus, we can deduce the real marking rate of each RED queue. These results are grouped and presented in the Tab \ref{tabcomplet}. They correspond to the roots computed for the flow \#1 and \#2 in the previous section \ref{reso}. We note that the observed average queue values have a low standard deviation. These values are almost constant for all the simulation.
  
\begin{table}[h]
	\begin{center}
	\begin{tabular}{|c|c|c|c|c|} 
		\hline         & Average   & Theoretical & \multicolumn{2}{|c|}{Estimated}\\
		        Queue  &  Size     & Marking     & \multicolumn{2}{|c|}{Marking Rate}\\
		               & (\# pkts) &  Rate       &  flow \#1    & flow \#2 \\  
		\hline Queue1 (E2--C1) & 55.5 & 11\% & 11\% & $\oslash$  \\ \cline{2-3}
		\hline Queue2 (C1--C2) & 60.5 & 21\% & 21\% & 20\% \\
		\hline Queue3 (C2--C3) & 72 & 44\% & 42\% & 44\% \\
		\hline Queue4 (C3--E3) & 77.5 & 55\% & 55\% & $\oslash$ \\
		\hline Queue5 (E1--C1) & 65.5 & 32\% & $\oslash$ & 31\% \\
		\hline Queue6 (C3--E4) & 87 & 74\% & $\oslash$ & 74 \% \\
		\hline
		\end{tabular}
	\caption{Average queue length and corresponding theoretical marking rate }
	\label{tabcomplet}
	\end{center}
\end{table}

These results globally correspond to the estimations with a slight difference explained by the size of the sample. Moreover, if we do a correlation between the results analytically obtained and those obtained by simulation in table \ref{tabcomplet}, we can notice that flows \#1 and \#2 estimate two marking rates in common corresponding to the two common routers crossed by both flows. Thus, not only these results correspond to the expected ones but they also underline an important aspect: it seems these measurements are not disturbed with each other and are perfectly independents (several other measurements, not presented here tend to confirm this fact). In other words, this allows to drive several measurements in parallel on a same network. We also verify, thanks to this simulation, that the hypothesis of network stability is sufficient. Thus, if we assume that the path used is relatively constant and the congestion level remains stable, this method allows a good estimation of the congestion level of the different routers of a given path.

\subsection{Convergence of this method}

\begin{figure}[htb!] 
		\subfigure[Evaluation of sigmas as a function of the number of received packets]{\includegraphics[width=4.3cm]{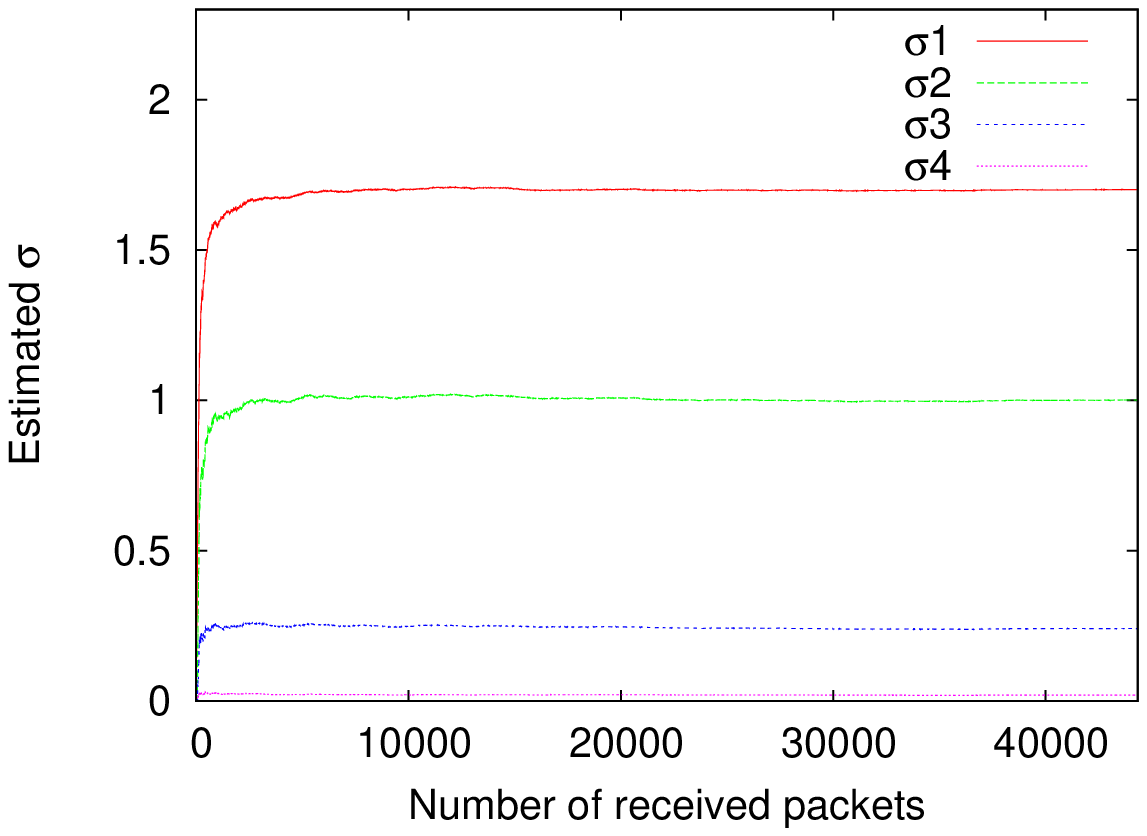}\label{fig:convergence1}}
		\subfigure[Computed marking rates as a function of the number of received packets]{\includegraphics[width=4.3cm]{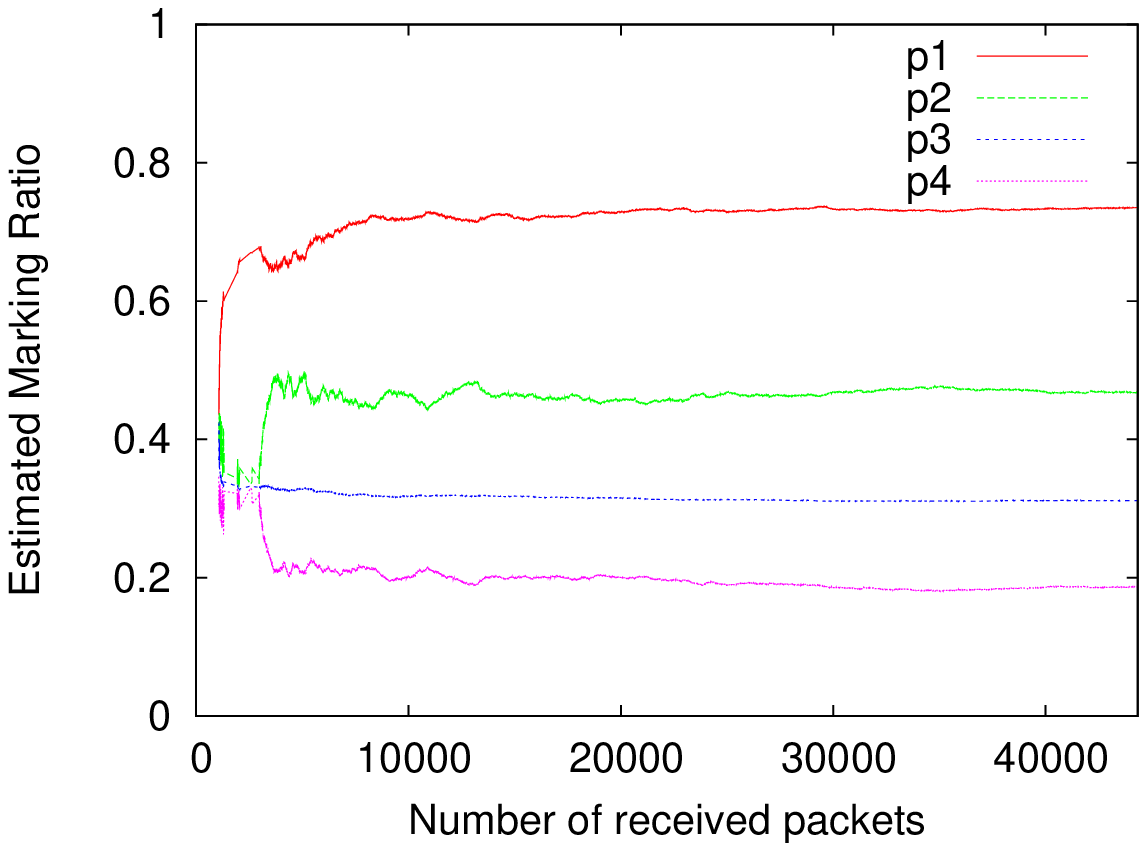}\label{fig:convergence2}}
		\caption{Evolution of sigmas and marking rates as a function of received packets}
	\end{figure}

As detailed previously, we adopt a probabilistic approach to solve this problem. We admittedly take in consideration the measurement uncertainty by solving $P(x) = \epsilon$. Nevertheless, it is necessary to focus on the convergence time of this solution. It means to assess when the size of the sample is big enough to correctly determine the different marking rates. To do so, we evaluate the different $\sigma$, directly linked to the coefficients of the solving polynomial every $50$ received packets. The evolution of these coefficients as a function of the number of received packets allow us to determine a threshold from which the value of these coefficients does not evolve anymore. A second threshold can also be set: the one which corresponds to the number of packets from which we can find the solutions to the equation $P(x) = \epsilon$. This approach is presented in figure \ref{fig:convergence1}.


As we can see in the Figure \ref{fig:convergence2}, we only need 4000 packets to have a correct estimation of the coefficients and about 8000 packets to reach a perfect estimation (equivalent to a 90 seconds transfer in our simulation) with an $\epsilon \pm 10^{-3}$. If we focus on Figure \ref{fig:convergence1}, we can note that between 3000 and 4000 received packets, the coefficients of the polynomial do not evolve much more. This underlines the accuracy necessary to establish the good solving polynomial. Indeed, we have to accurately estimate the $p(M_k^n)$ to have good results. Other simulations, not presented here, done over a similar topology but with routers less congested, have shown that these thresholds are slightly higher. In fact, the lower is the event corresponding to the marking of a packet, the higher the size of the sample has to be in order to observe this event and so to accurately estimate it. Respectively, the higher is the marking rates (equivalent to an important congestion) the smaller can be the size of the sample.

 	
\section{Conclusion}

In this article, we have proposed to increase the level of congestion information returned by TCP feedback messages with an ECN* marking scheme. ECN* enables the ECN field to count how many times a packet has crossed a congested router.
We define an algorithm able to estimate the congestion level of each queue of a given path through the analysis of the data collected. These preliminary results suggest that this method is reliable and robust to cross traffics.
In this study, we demonstrate the existing relationship between this ECN* marking rate and the filling level of each routers' queue.

However, several others investigations need to be driven in the context of dynamic networks and concerning the size of the sample statistic set. This method is not complete and we are currently investigating an extension of this algorithm robust to the dynamic changing of the 
network, based on a novel way to determine the areas of root, to allow a faster convergence to the solution.
Finally the ultimate step is obviously to determine how to interact with TCP congestion control and how TCP should take into account this new congestion feedback information. 

\bibliographystyle{plain}
\bibliography{biblio}

\begin{thebibliography}{10}

\bibitem{re-ecn}
B.~Briscoe, A.~Jacquet, T.~Moncaster, and A.~Smith.
\newblock {Re-ECN}: {A}dding accountability for causing congestion to {TCP/IP}.
\newblock Internet draft, Internet Engineering Task Force, July 2008.

\bibitem{floyd94ecn}
S.~Floyd.
\newblock {TCP} and explicit congestion notification.
\newblock {\em ACM Comp. Comm. Review}, 24(5):10--23, 1994.

\bibitem{rfc3649}
S.~Floyd.
\newblock {HighSpeed {TCP} for Large Congestion Windows}, December 2003.
\newblock Request for Comments 3649.

\bibitem{red}
S.~Floyd and V.~Jacobson.
\newblock Random early detection gateways for congestion avoidance.
\newblock {\em IEEE/ACM Transaction on Networking}, 1(4):397--413, August 1993.

\bibitem{xcp}
D.~Katabi, M.~Handley, and C.~Rohrs.
\newblock Congestion control for high bandwidth-delay product networks.
\newblock {\em SIGCOMM Comput. Commun. Rev.}, 32(4):89--102, 2002.

\bibitem{power}
A.~Kuzmanovic.
\newblock The power of explicit congestion notification.
\newblock {\em SIGCOMM Comput. Commun. Rev.}, 35(4):61--72, 2005.

\bibitem{le03effect}
L.~Le, J.~Aikat, K.~Jeffay, and F.~Smith.
\newblock The effects of active queue management on web performance.
\newblock In {\em In Proceedings of ACM SIGCOMM}, Karlsruhe, Germany, August
  2003.

\bibitem{medina05measuring}
A.~Medina, M.~Allman, and S.~Floyd.
\newblock Measuring the evolution of transport protocols in the internet.
\newblock {\em Computer Communication Review}, 35(2), April 2005.

\bibitem{conex}
T.~Moncaster, L.~Krug, M.~Menth, S.~Blake, and R.~Woundy.
\newblock The need for congestion exposure in the internet.
\newblock Internet draft, Internet Engineering Task Force, October 2009.

\bibitem{bmcc}
I.~Qazi, L.~Andrew, and T.~Znati.
\newblock Congestion control using efficient explicit feedback.
\newblock In {\em Proc. IEEE INFOCOM}, Rio de Janeiro, Brazil, 20-25 Apr 2009.

\bibitem{rfc2481}
K.~Ramakrishnan and S.~Floyd.
\newblock {A Proposal to add Explicit Congestion Notification (ECN) to IP},
  January 1999.
\newblock Request for Comments 2481.

\bibitem{stewart09}
L.~Stewart, G.~Armitage, and A.~Huebner.
\newblock Collateral damage: The impact of optimised {TCP} variants on
  real-time traffic latency in consumer broadband environments.
\newblock In {\em IFIP Networking 2009}. Springer, 2009.

\end{thebibliography}

\section*{Appendix}
\scriptsize
To demonstrate (\ref{eq:generale}), we use a proof by mathematical induction. The induction is done on the number of congested routers: $n$.

\emph{Basis:} the formula is demonstrated in part \ref{deuxrouteurs}. 

\emph{Inductive step:} $p(M_{k}^{n+1})$ is the probability for a packet to be marked $k$ times over a path of $n+1$ routers. The event $M_{k}^{n+1}$ can be decomposed. Indeed, be marked $k$ times over a path of $n+1$ routers is similar to be marked $k$ times by the $n$ first routers and not be marked by the router $n+1$; or to be marked $k-1$ times by the $n$ first routers and be marked by the router $n+1$. In terms of probability, this decomposition can be written as follows: 

	\begin{equation}
		\forall k, \ 1\leqslant k\leqslant n, \ 
		 p(M_{k}^{n+1})= p(M_{k}^{n}) (1-p_{n+1}) + p(M_{k-1}^{n}) p_{n+1} 
		\label{eq1}
	\end{equation}

Moreover, we have the following relations: 

	\begin{eqnarray}
		\forall k, \ 1\leqslant k\leqslant n, & \sigma_{k}^{n+1} & = \sigma_{k}^{n} + x_{n+1}. \sigma_{k-1}^{n} \nonumber \\
				& \sigma_{n+1}^{n+1} & = x_{n+1} \sigma_{n}^{n}
		\label{eq2}
	\end{eqnarray}

Developing (\ref{eq1}) and using (\ref{eq2}) we have : 
	\begin{align*}
		\forall k, \ 1\leqslant k\leqslant n,
		 p(M_{k}^{n+1}) & =  [\sum_{i=0}^{n-k} (-1)^{i} \begin{pmatrix} i+k \\ i \end{pmatrix} \sigma_{i+k}^{n}] (1-p_{n+1}) \\
		 & \ \ \ \ \ \ \ + [\sum_{i=0}^{n-k+1} (-1)^{i} \begin{pmatrix} i+k-1 \\ i \end{pmatrix} \sigma_{i+k-1}^{n}] p_{n+1} \\
		  & =  \sum_{i=0}^{n-k+1} (-1)^{i} \begin{pmatrix} i+k \\ i \end{pmatrix} \sigma_{i+k}^{n+1} 
	\end{align*}
The formula is so demonstrated for $n+1$. Then, we have : 

	\begin{eqnarray}
		 \forall k, \ 1\leqslant k\leqslant n+1, \ p(M_{k}^{n+1}) = \sum_{i=0}^{n+1-k} (-1)^{i} \begin{pmatrix} i+k \\ i \end{pmatrix}\sigma_{i+k}^{n+1} \notag 
	\end{eqnarray}
\begin{center}
 QED
\end{center}
\normalsize
\end{document}